%% file: main.tex
\documentclass[10pt,twocolumn,letterpaper]{article}
\usepackage[accsupp]{axessibility}
\usepackage{utils/cvpr}      

\usepackage{graphicx}
\usepackage{amsmath}
\usepackage{amssymb}
\usepackage{booktabs}
\usepackage{booktabs}
\usepackage{multirow}
\DeclareMathOperator*{\argmax}{argmax}
\usepackage{array}

\usepackage[pagebackref,breaklinks,colorlinks]{hyperref}

\usepackage[capitalize]{cleveref}
\crefname{section}{Sec.}{Secs.}
\Crefname{section}{Section}{Sections}
\Crefname{table}{Table}{Tables}
\crefname{table}{Tab.}{Tabs.}

\def\cvprPaperID{6189} 
\def\confName{CVPR}
\def\confYear{2022}

\begin{document}

\title{  Incremental Cross-view Mutual Distillation for \\
Self-supervised Medical CT Synthesis}

\author{Chaowei Fang$^1$\footnotemark[1]\quad Liang Wang$^1$\footnotemark[1]\quad Dingwen Zhang$^{2,3}$\footnotemark[2] \quad  Jun Xu$^4$ \quad Yixuan Yuan$^5$ \quad Junwei Han$^{2,3}$ 
\\
$^1$Xidian University\quad $^2$Northwestern Polytechnical University \quad \\ $^3$Hefei Comprehensive National Science Center  \quad $^4$Nankai University\quad $^5$City University of Hong Kong
}
\maketitle
\renewcommand{\thefootnote}{\fnsymbol{footnote}}
\footnotetext[1]{Equal contribution.}
\footnotetext[2]{Corresponding author.}
    
\input{sections/abstract}
\input{sections/introduction}
\input{sections/relatedwork}
\input{sections/method}
\input{sections/exper}
\input{sections/conclusion}

\vspace{1mm}
\noindent \textbf{Acknowledgement.} This work was supported in part by Key-Area Research and Development Program of Guangdong Province (No. 2021B0101200001), in part by  the National Natural Science Foundation
of China (No. 62003256, 61876140, 62027813, U1801265, and U21B2048), in part by Open Research Projects of Zhejiang Lab (No. 2019kD0AD01/010), and in part by MindSpore which is a new deep learning computing framework\footnote{https://www.mindspore.cn/}.

{\small
\bibliographystyle{ieee_fullname}
\bibliography{egbib}
}

\end{document}

%% file: sections/abstract.tex
\begin{abstract}
Due to the constraints of the imaging device and high cost in operation time, computer tomography (CT) scans are usually acquired with low within-slice resolution. Improving the inter-slice resolution is beneficial to the disease diagnosis for both human experts and computer-aided systems. To this end, this paper builds a novel medical slice synthesis to increase the inter-slice resolution. Considering that the ground-truth intermediate medical slices are always absent in clinical practice, we introduce the incremental cross-view mutual distillation strategy to accomplish this task in the self-supervised learning manner. Specifically, we model this problem from three different views: slice-wise interpolation from axial view and pixel-wise interpolation from coronal and sagittal views. Under this circumstance, the models learned from different views can distill valuable knowledge to guide the learning processes of each other. We can repeat this process to make the models synthesize intermediate slice data with increasing between-slice resolution. To demonstrate the effectiveness of the proposed approach, we conduct comprehensive experiments on a large-scale CT dataset. Quantitative and qualitative comparison results show that our method outperforms state-of-the-art algorithms by clear margins.
\end{abstract}

%% file: sections/introduction.tex
\input{figures/fig-teaser}

\input{figures/fig-distill3views}
\vspace{-2mm}
\section{Introduction}
High-resolution CT volume data can provide high-quality detail for organs and tissues, thus are valuable for computer-aided diagnosis. However, due to the constraints of the imaging device, the between-slice resolution of the acquired CT volume is not sufficiently high in practical clinical scenarios, which makes these volume data hard to provide the desired imaging detail for the disease diagnosis.

To solve this problem, a novel task, called medical slice synthesis, has been arising recently. The goal is to synthesize intermediate imagery content between original adjacent slices. Peng \textit{et al.},~\cite{peng2020saint} made the earliest attempt by implementing pixel-wise interpolation processes on the coronal-view and sagittal-view images and then fusing the results interpolated from two views. However, this method requires large-scaled ground-truth training data, which we cannot conveniently acquire in practice. 

This paper explores a self-supervised learning framework to train the slice synthesizer without the ground-truth data. Specifically, we find that another under-explored way is to formulate it as a slice-wise interpolation problem for the axial-view images (See Fig.~\ref{fig:teaser}). Namely, intermediate slices can be inferred from the context information of two adjacent slices in the axial view. Since pixel-wise and slice-wise interpolation modeling tries to synthesize the missing detail by exploring different kinds of spatial context, the two modeling processes tend to capture helpful yet distinct patterns towards the same ultimate goal. Thus, we can jointly use the two modeling processes to address the medical slice synthesis problem and collaborate them to provide complementary knowledge for each other. Each interpolation model can be learned under the guidance of the other ones, thus avoiding the requirement of ground-truth training data.

We propose an incremental cross-view mutual distillation pipeline for training medical CT slice synthesis models to take advantage of slice synthesis algorithms from multiple views.
Considering that structural information appears to have different characteristics across views and models learned from different views have their superiority (see Fig.~\ref{fig:distill}), we involve three modeling components in the learning process: 1) slice-wise interpolation in axial view; 2) pixel-wise interpolation in coronal view; 3) pixel-wise interpolation in sagittal view. 
We set up a U-shape network with memorization capacity to implement the slice-wise interpolation and adopt an existing image super-resolution network~\cite{niu2020single} to achieve pixel-wise interpolation. 

To lean such deep models, we propose a two-stage learning framework. In the first learning stage, we downsample the resolution of original volumes and then use the downsampled and original volume data to learn single-view slice synthesis models. To enable the model to upscale the resolution of the original volume data without any external supervision, we further design a cross-view mutual distillation process in the second learning stage. We constrain the pairs of predictions on the original volume data produced by axial-view slice-wise interpolation and coronal/sagittal-view pixel-wise interpolation models. An illustration of our proposed method is presented in Fig.~\ref{fig:teaser}. The knowledge distillation mechanism enables the slice-wise and pixel-wise interpolation models to learn from each other and fuse the advantages of different image recovery models learned from different perspectives. Finally, we incrementally increase the between-slice resolution from the three perspectives and apply the cross-view mutual distillation on predictions with very high resolution, enhancing the knowledge exchange across views in self-supervised slice synthesis.

The main contributions of this paper are as follows.
\begin{itemize}
\item A pioneering effort is made to implement the self-supervised CT slide synthesis, modeling slice-wise interpolation for the axial view and pixel-wise interpolation for the coronal and sagittal views.
\item A novel self-supervised learning framework is established, based on single-view internal learning and incremental cross-view mutual distillation. 
\item Extensive experiments on a CT collection (composed of three existing CT datasets) demonstrate that our proposed method achieves state-of-the-art performance. 
\end{itemize}


%% file: figures/fig-teaser.tex
\begin{figure}[t]
\centering
\includegraphics[width=1\linewidth]{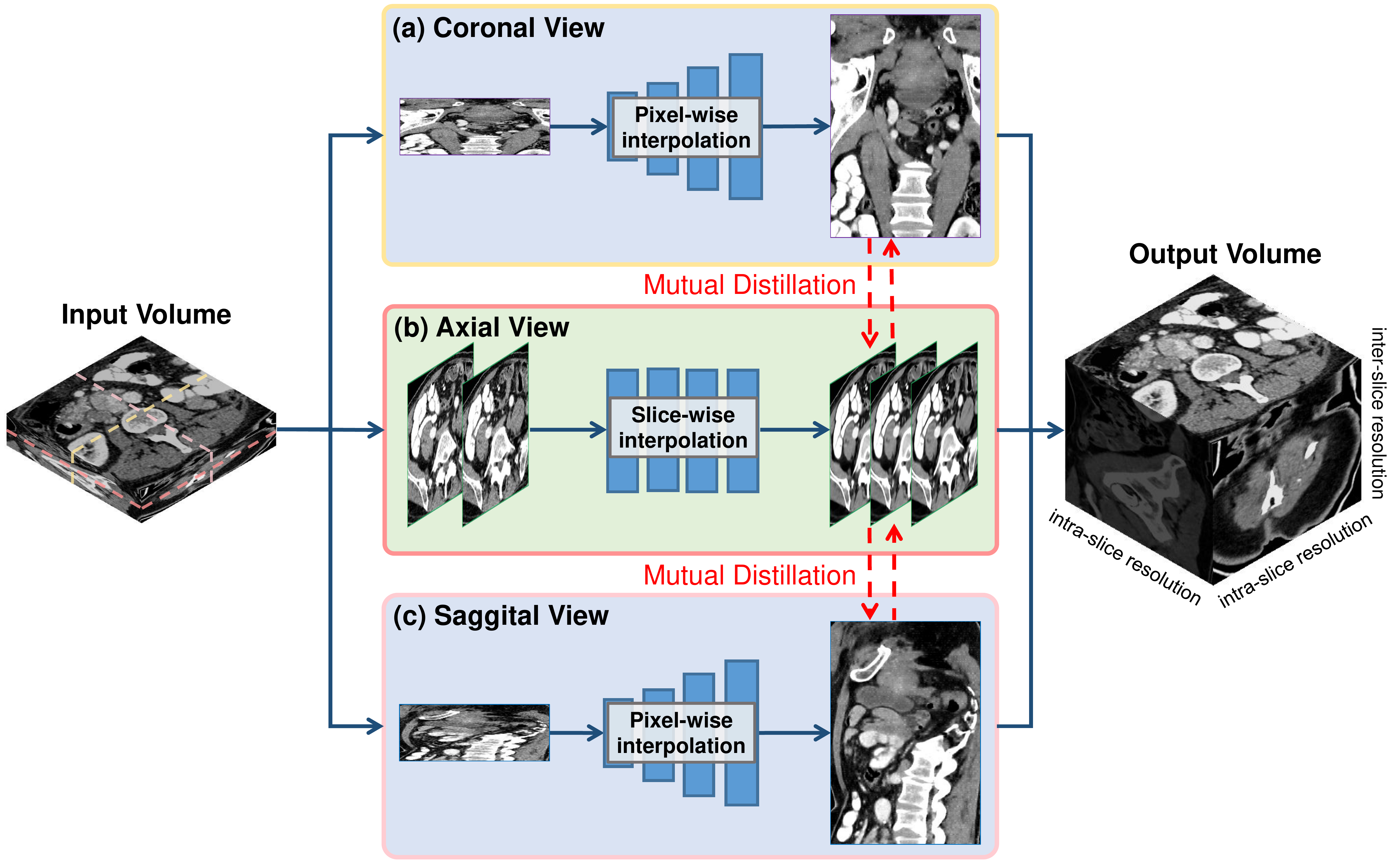}\vspace{-3mm}
\caption{ Pixel-wise interpolation in coronal and sagittal views, and slice-wise interpolation in  axial view can increase the inter-slice resolution of the input volume individually. We propose a cross-view knowledge distillation framework to settle the self-supervised CT slice synthesis task.}\label{fig:teaser} \vspace{-5mm}
\end{figure}

%% file: figures/fig-distill3views.tex
\begin{figure*}[t]
\centering
\includegraphics[width=1\linewidth]{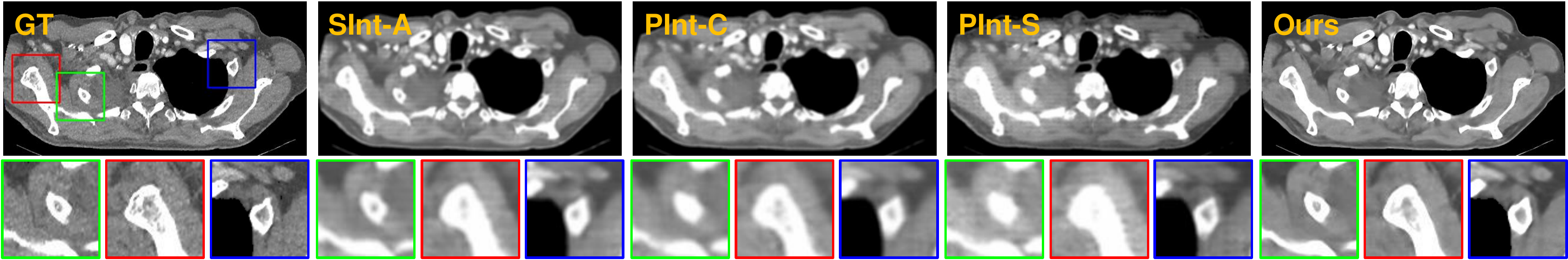}\vspace{-3mm}
\caption{ Slice-wise interpolation in  axial view (SInt-A), and pixel-wise interpoloatin in coronal (PInt-C) and sagittal (PInt-S) views, have their own superiority in synthesizing inter-slice images. Our proposed cross-view mutual distillation can combine the learned knowledge from three types of interpolation algorithms.}\label{fig:distill}
\vspace{-5mm}
\end{figure*}

%% file: sections/relatedwork.tex
\section{Related Work}

\noindent \textbf{Medical slice synthesis} is targeted at hallucinating inter-slice detail which is critical to high-level disease diagnosis for both radiologists and computer-based intelligent systems.
Recently, 3D neural networks~\cite{chen2018efficient,sanchez2018brain,you2019ct,tang2021self,hatamizadeh2022unetr} are extensively applied in processing and understanding medical volumetric data. 
The main drawback of using 3D neural networks is the huge amount of network parameters and memory consumption.
SAINT~\cite{peng2020saint} is a two-stage framework to solve the slice synthesis task.
It first employs 2D convolutional neural networks (CNNs) to enlarge sagittal and coronal images individually, and then fuse the enlarged images of two views to produce the final result.

Learning slice synthesis CNNs requires a large number of paired LR and HR volumes.
However, HR volumes are usually not available in practical medical scenarios. 
Thus, it is essential to develop unsupervised optimization algorithms for medical slice synthesis CNNs.
As far as we know, few work is devoted to addressing this task. 
In this paper, we focus on the unsupervised slice synthesis task, and propose a cross-view mutual distillation pipeline , twisting slice-wise interpolation in axial view and pixel-wise interpolation in coronal and sagittal views.

\input{figures/fig-pipeline}
\vspace{1mm}
\noindent \textbf{Video Frame Interpolation}.
Slice-wise interpolation is highly related to video frame interpolation. In videos, the differences between consecutive frames are mainly caused by object or camera motions. Thus, video interpolation algorithms usually rely on optical flow fields~\cite{niklaus2018context,peleg2019net,li2020video,Choi_2021_ICCV,Park_2021_ICCV}, adaptive kernels~\cite{niklaus2017video,niklaus2017videosep}, or flow-based adaptive kernels~\cite{lee2020adacof} to interpolate intermediate transition frames from temporally neighboring frames.
Aiming at tackling frame interpolation under complex motions and severe occlusions,
\cite{kalluri2020flavr,choi2020channel,xu2021TMNet} adopts an image reconstruction pipeline without using motion fields and adaptive kernels which are difficult to be estimated when there exist large motions and severe occlusions in the input video.
The slice synthesis task is more challenging since the different slices contains totally different content and there exist no explicit correspondence relations between adjacent slices. 

\vspace{1mm}
\noindent \textbf{Image Super-Resolution} (SR).
As a fundamental and long-lasting topic in image processing, super-resolution attracts lots of research attention.
Dong \textit{et al.}~\cite{dong2015image} apply convolutional neural networks in image super-resolution for the first time. 
Mainstream SR methods depend on various CNN backbones~\cite{dong2015image,ledig2017photo,zhang2018residual,Xie_2021_ICCV,Wang_2021_CVPR}.
MetaSR~\cite{hu2019meta} proposes to tackle the SR task of arbitrary scales through dynamic kernels learned from the pixel coordinates and upscaling factor.
HAN~\cite{niu2020single} introduces the holistic attention to explore cross-position, cross-channel and cross-layer dependencies for promoting SR performance.
The slice synthesis can be implemented via image SR in the coronal and sagittal views.


\vspace{1mm}
\noindent \textbf{Knowledge Distillation}.
The concept of knowledge distillation is first proposed for model compression in~\cite{bucilua2006model}. Hinton \textit{et al.}~\cite{hinton2015distilling} define knowledge distillation as the task of transferring the knowledge of a teacher model which can be a very large model or an ensemble of multiple models to a student model. They also propose a distillation strategy through using the soft outputs of the teacher model to guide the training of the student model.
Henceforth, a lot of literature focuses on devising more effective distillation algorithms~\cite{romero2014fitnets,zagoruyko2016paying,mirzadeh2020improved}. 
Our proposed method is most related to the mutual learning~\cite{zhang2018deep}, in which an ensemble of student models learn from each other.
The major difference of our method to mutual learning is that, the student networks in our method are constructed from different views of the volumetric data and devised for addressing different tasks, namely slice-wise or pixel-wise interpolation.



%% file: figures/fig-pipeline.tex
\begin{figure*}[t]
\centering
\includegraphics[width=0.96\linewidth]{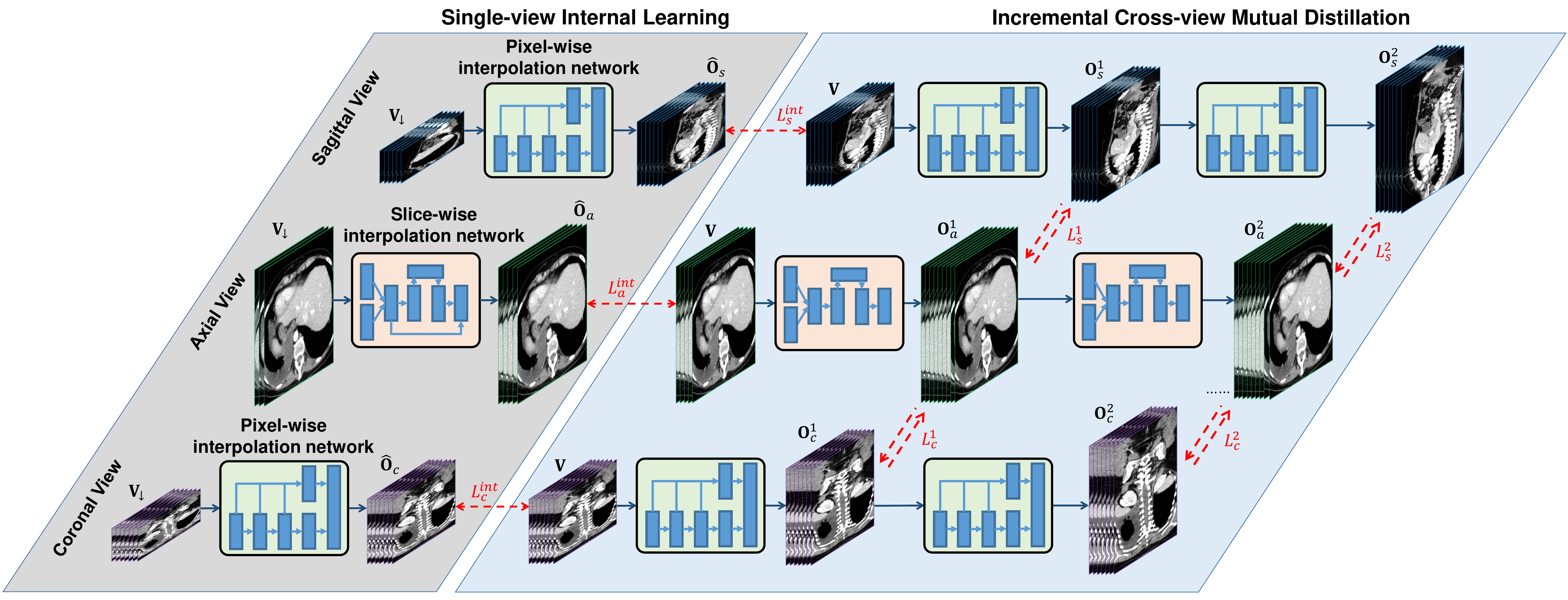} \vspace{-3mm}
\caption{ Overall pipeline of our method. First, internal learning is used to regularize single-view interpolation models via regarding downsampled and original volumes as training samples. Then, an incremental cross-view mutual distillation pipeline is devised for knowledge exchange between the slice-wise interpolation in the axial view and the pixel-wise interpolation in the coronal and sagittal views.  }\label{fig:pipeline}
\vspace{-5mm}
\end{figure*}

%% file: sections/method.tex
\input{figures/fig-internet}

\section{Proposed Method}

Given a 3D volume $\mathbf V\in \mathbb R^{h\times w\times l}$, we assume that $r-1$ ($r\geq2$) slices should be interpolated between every two consecutive slices. This means that a volume defined by $\mathbf O\in \mathbb R^{h\times w\times (rl-r+1)}$ is expected to be produced. $\mathbf V$ can be decomposed into 2D images in the axial, coronal and saggital views, yielding $\{\mathbf X_a^i\in \mathbb R^{h\times w}\}_{i=1}^l$, $\{\mathbf X_c^j\in \mathbb R^{w\times l}\}_{j=1}^h$, and $\{\mathbf X_s^k\in \mathbb R^{h\times l}\}_{k=1}^w$, respectively. 
We can achieve the goal with three models that perform slice-wise interpolation in the axial view and pixel-wise interpolation in the coronal and sagittal views. The concrete model design can be referred to in Sec. \ref{Interpolation Models}.

Since actual training data is hard to
obtain, we follow the degradation operation in~\cite{peng2020saint} or \cite{jiang2018super} to approximate the real downsampling case. Under this circumstance, single-view internal learning is first used to constrain the three models with the help of down-sampled volumes.
Then, the slice-wise and pixel-wise interpolation models are constrained via the consistency between volume data enlarged by them for knowledge distillation across views.
The overall framework of our method is presented in Figure~\ref{fig:pipeline}.
Though 3D convolution can be alternatively used, we implement our framework with 2D convolution-based modules in consideration of computational efficiency.


\subsection{Interpolation Models}
\label{Interpolation Models}
\subsubsection{Slice-wise Interpolation}

The slice synthesis can be implemented via inserting intermediate slices between every two adjacent slices. Inspired from~\cite{choi2020channel}, we build up a CNN model for slice-wise interpolation in the axial view (see Fig.~\ref{fig:interpnet}).
Given two consecutive slices $\mathbf X_a^i\in \mathbb R^{h\times w}$ and $\mathbf X_a^{i+1} \in \mathbb R^{h\times w}$, a convolution layer with the kernel size of $3\times3$ and the dimension of 3 is used to extract two preliminary feature maps. They are rearranged into tensors $\mathbf F^i \in \mathbb R^{\frac{h}{8} \times \frac{w}{8} \times 192}$ and $\mathbf F^{i+1} \in \mathbb R^{\frac{h}{8} \times \frac{w}{8} \times 192}$ through the space-to-depth transformation operation. Then, a U-shape architecture is devised to fully explore multiple features of different layers  to estimate the intermediate slices between $\mathbf X^i_a$ and $\mathbf X^{i+1}_a$, namely $\{\mathbf Y^{(i-1)r+t}_a\}_{t=2}^r$. 

$\mathbf F^i$ and $\mathbf F^{i+1}$ are concatenated and then compressed into a tensor $\mathbf E_0^i\in\mathbb R^{\frac{h}{8} \times \frac{w}{8} \times 192}$ via a $3\times3$ convolution layer. Then, three groups of residual blocks with channel attentions~\cite{choi2020channel} are used to produce multiple feature maps $\mathbf E_1^i$, $\mathbf E_2^i$ and $\tilde{\mathbf E}_3^i$. 
Each group is composed of 12 residual blocks.
$\tilde{\mathbf E}_3^i$ is added to $\mathbf E_0^i$ through a skip connection, and another $3\times3$ convolution is attached to produce the final feature map $\mathbf E_3^i$.

Considering CT images usually share high similarities (e.g., anatomical structures) across persons,  we incorporate a memory bank~\cite{park2020learning} $\mathbf M\in \mathbb R^{m\times d}$ to store the common patterns. $m$ is the number of items in the memory bank. 
All points in $\mathbf E_3^i$ are reconstructed with $\mathbf M$, deriving a new feature map $\mathbf D_3^i$. The  linear combination of items in $\mathbf M$ is used to infer every point in $\mathbf D_3^i$,
\begin{equation}
\mathbf D_3^i[x,y] = \sum_{z=1}^m p^i_{x,y,z} \mathbf M[z],
\end{equation}
where $\mathbf D^i_3[x,y]$ represents the feature vector at position $(x,y)$ of $\mathbf D^i_3$, and $\mathbf M[z] $ indicates the $z$-th item of the memory bank $\mathbf M$. $p^i_{x,y,z}$ indicates the weight coefficient between $\mathbf E^i_3[x,y]$ and $\mathbf M[z]$,
\begin{equation}
p^i_{x,y,z} = \frac{\textrm{exp}(\mathbf E^i_3[x,y]\circ \mathbf M[z])}{\sum\limits_{z^\prime=1}^{m} \textrm{exp}(\mathbf E^i_3[x,y]\circ \mathbf M[z^\prime])}.
\end{equation}
Here, `$\circ$' indicates the inner product operation.
During the training stage, the memory bank is continuously updated through accumulating the emerging patterns in $\mathbf E^i_3$.
\begin{align}
q^i_{x,y,z} &= \frac{\textrm{exp}(\mathbf M[z] \circ \mathbf E^i_3[x,y])}{\sum\limits_{(x^\prime,y^\prime)\in\mathcal U_k} \textrm{exp}(\mathbf M[z] \circ \mathbf E^i_3[x^\prime,y^\prime])}, \\
q^i_{x,y,z} &\leftarrow q^i_{x,y,z}/\max\limits_{(x^\prime,y^\prime)\in\mathcal U_k} q^i_{x^\prime,y^\prime,z}, \\
\mathbf M[z]& \leftarrow  \mathbf M[z]+\sum_{(x^\prime,y^\prime)\in\mathcal U_k} q^i_{x^\prime,y^\prime,z}\mathbf E^i_3[x^\prime,y^\prime], \\
\mathbf M[z]& \leftarrow  \mathbf M[z]/||\mathbf M[z]||_2.
\end{align}
$\mathcal U_k$ represents the set of points whose nearest neighbor in the memory bank is $\mathbf M[k]$. Based on the above process, the memory bank is updated by accumulating patterns across all training CT slices. It can store representative visual patterns for CT slice reconstruction. 

The decoding stage is constituted by three consecutive modules. Each module contains one $3\times3$ convolution layer and twelve residual blocks. $\mathbf D^i_3$ is regarded as the input of the first stage. Skip connections are used to propagate $\mathbf E^i_2$ and $\mathbf E^i_1$ into the second and third stages of the decoder, respectively. Finally, a $3\times3$ convolution followed by a depth-to-space operation is employed to produce intermediate slices.
By means of the above slice interpolation model, the input volume is interpolated into a new volume with more slices, $\{\mathbf Y_a^i\}_{i=1}^{rl-r+1}$, where $\mathbf Y_a^i=\mathbf X_a^{(i-1)\%r}$, if $i\%r=1$. We denote the interpolated volume as $\mathbf O_a=\textrm{SInt}_a(\mathbf V|\Theta_s)$. $\Theta_s$ denotes the parameters of the interpolation model. 


\subsubsection{Pixel-wise Interpolation}
The other perspective for slice synthesis is the pixel-wise interpolation, based on the super-resolving of the images in the coronal or sagittal view. We use the image super-resolution network proposed in~\cite{niu2020single} for pixel-wise interpolation. The coronal and sagittal views share the same model. The coronal images $\{\mathbf X_c^j\}_{j=1}^h$ and sagittal images $\{\mathbf X_s^k\}_{k=1}^w$ are super-resolved by a factor of $r$ along the longitudinal axis, resulting in $\{\mathbf Y_c^j\}_{j=1}^h$ and $\{\mathbf Y_s^k\}_{k=1}^w$ respectively.  
The last $r-1$ columns of super-resolved images are abandoned to make the shape consistent with the volume produced by the slice-wise interpolation model. 
These super-resolved coronal and sagittal images can be stacked into new volumes $\mathbf O_c$ and $\mathbf O_s$ respectively. We denotes the pixel-wise interpolation processes in the coronal and sagittal view as, $\mathbf O_c=\textrm{PInt}_c(\mathbf V|\Theta_p)$ and $\mathbf O_s=\textrm{PInt}_s(\mathbf V|\Theta_p)$ respectively. $\Theta_p$ denotes parameters of the pixel-wise interpolation model.
\vspace{1mm}

During the inference phase, the final interpolation result $\mathbf O$ is obtained via fusing $\mathbf O_a$, $\mathbf O_c$ and $\mathbf O_s$,
\begin{equation}
    \mathbf O[x,y,z]=\begin{cases}
     \frac{\mathbf O_c[x,y,z]+\mathbf O_s[x,y,z]}{2} & \textrm{if}\;z\%r=1  \\
    \frac{\mathbf O_a[x,y,z]+\mathbf O_c[x,y,z]+\mathbf O_s[x,y,z]}{3} & \textrm{else}
    \end{cases}
\end{equation}

\subsection{Learning Procedure}
\label{Learning Procedure}
\subsubsection{Single-view Internal Learning}
An internal learning strategy is adopted to optimize individual single-view slice-wise or pixel-wise interpolation models. The original volume is down-sampled by the factor of $r$ along the axial view, resulting in $\mathbf V_\downarrow \in \mathbb R^{h\times w\times \lfloor \frac{l}{r} \rfloor}$. Feeding $\mathbf V_\downarrow$ into the slice-wise and pixel-wise interpolation models, we can obtain upsampled volumes: $\hat{\mathbf O}_a=\textrm{SInt}_a(\mathbf V_\downarrow)$, $\hat{\mathbf O}_c=\textrm{PInt}_c(\mathbf V_\downarrow)$, and $\hat{\mathbf O}_s=\textrm{PInt}_s(\mathbf V_\downarrow)$. Here, parameters are neglected for briefness.

Regarding the original volume as the ground-truth, we calculate the training loss with the mean square error (MSE) function. Besides, to strengthen the restoration on high-frequency details, we extract three scales of wavelet coefficients and use MSE to constrain the distances on the LH (horiz), HL (vertic), and HH (diag) coefficients of each wavelet decomposition scale. The overall loss functions used in the single-view internal learning are as follows.
\begin{align}
L_a^{int}&=\textrm{MSE}(\hat{\mathbf O}_a,\mathbf V)+\sum_{t=1}^3  \textrm{MSE}(\textrm{WT}^{(t)}_a(\hat{\mathbf O}_a), \textrm{WT}^{(t)}_a(\mathbf{V})), \\
L_c^{int}&=\textrm{MSE}(\hat{\mathbf O}_c,\mathbf V)+\sum_{t=1}^3 
\textrm{MSE}(\textrm{WT}^{(t)}_c(\hat{\mathbf O}_c), \textrm{WT}^{(t)}_c(\mathbf{V})),\\
L_s^{int}&=\textrm{MSE}(\hat{\mathbf O}_s,\mathbf V)+\sum_{t=1}^3 \textrm{MSE}(\textrm{WT}^{(t)}_s(\hat{\mathbf O}_s), \textrm{WT}^{(t)}_s(\mathbf{V})).
\end{align}
$\textrm{WT}^{(t)}_a(\cdot)$, $\textrm{WT}^{(t)}_c(\cdot)$, and $\textrm{WT}^{(t)}_s(\cdot)$ calculates the $t$-th scale of wavelet coeffficients from the axial, coronal, and sagittal images of the input volume respectively. 
The restored volumes may have a smaller size than $\mathbf V$ due to the quantization effect, and excess voxels of $\mathbf V$ are neglected when calculating the above loss functions.

\subsubsection{Incremental Cross-view Mutual Distillation}
Given axial, coronal, and sagittal images originating from the same volume, the slice-wise and pixel-wise interpolation models have specific superiority in synthesizing details since different context is explored. We devise an MSE-based consistent constraint to make the two kinds of models teach each other so that the specific advantages of the three interpolation schemes are combined to promote the ultimate interpolation performance. Such a cross-view mutual distillation method can tackle the dilemma in which the ground-truth training data is absent. 
Practically, we repeat the slice-wise and pixel-wise interpolation for $n$ times, deriving of $\mathbf O_a^{n}=\textrm{SInt}_a^{(n)}(\mathbf V)$, $\mathbf O_c^{n}=\textrm{PInt}_c^{(n)}(\mathbf V)$, and $\mathbf O_s^{n}=\textrm{PInt}_s^{(n)}(\mathbf V)$.
The consistency constraints between the slice-wise interpolation result in axial view and the pixel-wise interpolation result in coronal/sagittal view are formulated as follows,
\begin{eqnarray}
\label{eq:loss-cc-ac} L_c^n&=\sum_{(x,y,z)\in \mathbb T^n_c(\gamma)} \frac{(\mathbf O_a^n[x,y,z]-\mathbf O_c^n[x,y,z])^2}{|\mathbb T^n_c(\gamma)|}, \\
\label{eq:loss-cc-as} L_s^n&=\sum_{(x,y,z)\in \mathbb T^n_s(\gamma)} \frac{(\mathbf O_a^n[x,y,z]-\mathbf O_s^n[x,y,z])^2}{|\mathbb T^n_s(\gamma)|},
\end{eqnarray}
where $\mathbb T^n_c(\gamma)$ ($\mathbb T^n_s(\gamma)$) denotes the set of $\gamma$ percents of points with smallest loss values between $\mathbf O_a^n$ and $\mathbf O_c^n$ ($\mathbf O_s^n$). Assume he largest number of interpolation times be $N$. The overall objective functions for the cross-view mutual distillation are formulated as, $L_c^{cmd} = \frac{1}{N}\sum_{n=1}^N L_c^n$, and $L_s^{cmd} = \frac{1}{N}\sum_{n=1}^N L_s^n$.

\subsubsection{Overall Objective Function}
Apart from the single-view internal learning and cross-view mutual distillation loss functions, the compactness ($L^{com}$) and separateness ($L^{sep}$) constraints as in~\cite{park2020learning}, are used to regularize the memory bank,
\begin{align}
    \nonumber L^{com}&=\sum_{i=1}^{l-1} \sum_{x=1}^{h/8} \sum_{y=1}^{w/8} \|\mathbf E^i_3[x,y]-\mathbf M[z^i_{\textrm{pos}}(x,y)]\|_2, \\ 
    & s.t.\;\; z^i_{\textrm{pos}}(x,y) =\argmax_{z^\prime} p^i_{x,y,z^\prime}; \\
    \nonumber L^{sep}&=\sum_{i=1}^{l-1} \sum_{x=1}^{h/8} \sum_{y=1}^{w/8} \max(\|\mathbf E^i_3[x,y]-\mathbf M[z^i_{\textrm{pos}}(x,y)]\|_2 \\
    \nonumber &-  \|\mathbf E^i_3[x,y]-\mathbf M[z^i_{\textrm{neg}}(x,y)]\|_2 + \alpha, 0), \\ 
    & s.t.\;\; z^i_{\textrm{neg}}(x,y) =\argmax_{ z^\prime\neq z^i_{\textrm{pos}}(x,y) } p^i_{x,y,z^\prime}.
\end{align}
$\alpha$($=1$) is a constant. The complete objective function is formed through summing up the above losses, $L=L_a^{int}+L_c^{int}+L_s^{int}+0.15*(L_c^{cmd}+L_s^{cmd})+0.1*(L^{com}+L^{sep})$. The weighting factors are chosen empirically. 

%% file: figures/fig-internet.tex
\begin{figure*}[t]
\centering
\includegraphics[width=0.96\linewidth]{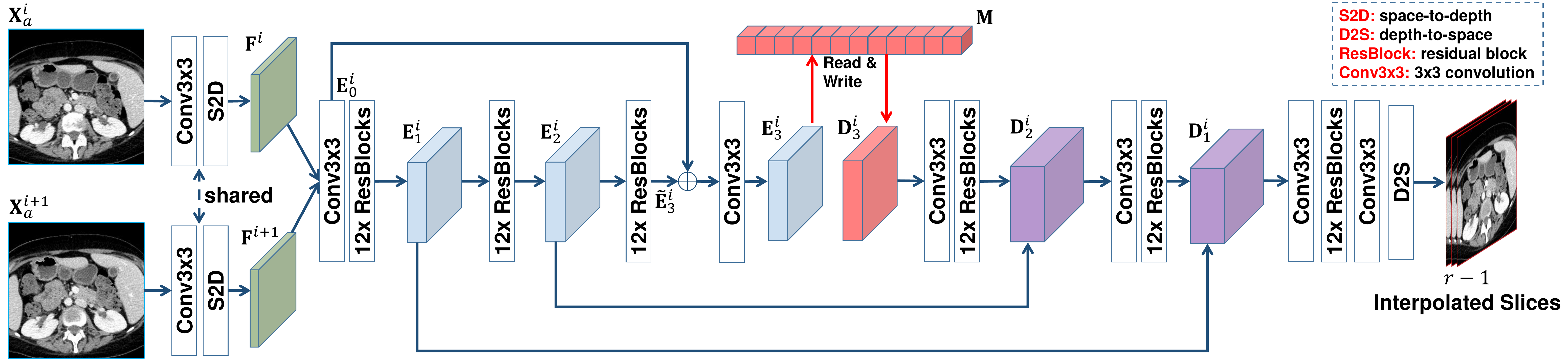} \vspace{-3mm}
\caption{ Network architecture of the slice-wise interpolation model. Given two adjacent slices, a U-shape network constituted by convolution layers, residual groups~\cite{choi2020channel} and a memory bank~\cite{park2020learning}, synthesizes $r-1$ intermediate slices.  }\label{fig:interpnet}
\vspace{-5mm}
\end{figure*}

%% file: sections/exper.tex
\input{figures/fig-comp}

\section{Experiments}
\subsection{Experimental Settings} \label{sec:exper-setting}
\noindent \textbf{Dataset.} The CT Dataset consists of 560 volumes, which are collected from the Medical Segmentation Decathlon challenge~\cite{simpson2019large}, including 131, 126, and 303 volumes for liver, colon and hepatic vessel segmentation, respectively. The spatial size is $512\times512$ and the number of slices is in the range of 24 to 917. 
The within-slice resolution ranges from 0.5mm to 1.0mm, and the between-slice resolution ranges from 0.7mm to 8.0mm. Fifty volumes with the thinnest slices are used for testing, and the other 510 volumes are used for training. All volumes are downsampled by the factor of $r$ in the axial view, while high-resolution volumes are only used for validating algorithm performance. Two degradation strategies are used for validating interpolation algorithms: 1) Low-resolution volumes are synthesized via directly sampling one slice every $r$ slices in the axial view; 2) Low-resolution volumes are generated by blurring and down-sampling. Then, Gaussian noises are used to distort the down-sampled volumes.

\noindent \textbf{Evaluation Metrics.} We use two metrics, including PSNR (Peak Signal-to-Noise Ratio) and SSIM (Structural Similarity Index). 
SSIM is calculated independently on axial, coronal, and sagittal images, denoted by SSIM$_a$, SSIM$_c$, and SSIM$_s$, respectively.

\noindent \textbf{Implementation Detail.} During training, we only use central 256*256 regions of CT slices, which are further decomposed into $128\times128$ patches.  Adam~\cite{kingma2014adam} is chosen for network optimization. The model is trained for 50 epochs with a batch size of 4. The learning rate is initially set to $10^{-4}$ and decayed by 0.1 after ten epochs. By default, $m$, $\gamma$, and $N$ is set to 10, 40\%, and 2, respectively. We test three cases for the upsampling factor $r$ (2, 3, and 4). Each training volume is composed of 7, 7, and 9 slices for $r=2$, $r=3$, and $r=4$, respectively.


\input{tables/tab-ct}
\input{tables/tab-ct-blurred}
\input{tables/tab-mri}
\subsection{Comparisons against Existing Methods}
In this section, we compare our method against pixel-wise interpolation algorithms (including RDN~\cite{zhang2018residual}, DPSR~\cite{zhang2019deep} and MetaSR~\cite{hu2019meta} which are originally devised for tackling image super-resolution), slice-wise interpolation methods (including RRIN~\cite{li2020video} and AdaCoF~\cite{lee2020adacof} which are originally proposed for settling video frame interpolation), and the slice interpolation method  SAINT~\cite{peng2020saint}. 
\vspace{1mm}

\noindent \textbf{Quantitative Comparisons.} Experimental results on the CT dataset are reported in Table~\ref{tab:ct} and~\ref{tab:ct-blurred}. Our proposed method outperforms all algorithms by clear margins on both degradation settings. 
For example, under the $4\times$ interpolation setting, our method achieves 41.11dB and 37.87dB PSNR, which are 2.69dB and 1.17dB higher than the scores of SAINT, on the two degradation strategies, respectively.
\vspace{1mm}

\noindent \textbf{Qualitative comparisons} of our method against existing methods are presented in Fig.~\ref{fig:comp}. We also visualize the super-resolution performance of SAINT and our method in coronal and sagittal views under the $4\times$ upsampling setting in Fig.~\ref{fig:real}. Our method has clearer structures and more apparent organ boundaries than other methods. 
\vspace{1mm}

\noindent \textbf{Model Size \& Time Cost.} For $4\times$ slice synthesis, the number of parameters of SAINT and our method is 44.2M and 46.9M, respectively. The training processes of SAINT and our method cost 18 and 31 hours, respectively. When processing a $512\times512\times36$ volume, SAINT and our method consume 35.96 and 13.25 seconds, respectively. 

\input{figures/fig-real.tex}
\input{tables/tab-ablation}

\subsection{Ablation Study}
This subsection conducts extensive inner comparisons on the CT dataset under the $4\times$ interpolation setting. Here, LR volumes are synthesized via direct downsampling. Core components of our method are teased apart to validate their effectiveness. The results are reported in Table \ref{tab:ablation}.  

\vspace{1mm}
\noindent \textbf{Efficacy of cross-view mutual distillation}
is validated by removing consistency constraints $L_c^{cmd}$ or $L_s^{cmd}$. In the baseline method, both $L_c^{cmd}$ and $L_s^{cmd}$ are not used, which means the cross-view mutual distillation is not applied. 
Compared to the baseline method, the full version of our approach brings PSNR and SSIM$_\textrm{a}$ gain of 2.53dB and 0.0118, respectively.
Since pixel-wise interpolation in the coronal and sagittal views explore different context information for increasing the between-slice resolution, the knowledge learned from the two views is complementary to each other.
Without distillation between axial view and coronal/sagittal view, the PSNR is decreased by 0.85dB/0.87dB in contrast to the PSNR of the full version.
The distillation from two views performs better than the distillation with the single coronal or sagittal view. This can also be observed from an example of qualitative comparison in Fig. \ref{fig:inner}.
\input{figures/fig-inner}

\input{tables/tab-ensemble}
\vspace{1mm}
\noindent \textbf{Efficacy of Incremental Interpolation.}
As shown in Table \ref{tab:ablation}, using interpolation only once ($N=1$) increases PSNR and SSIM$_\textrm{a}$ by 1.98dB and 0.0183, respectively. Applying two interpolation times ($N=2$) can further improve the result with 0.64dB higher PSNR, compared to the variant with $N=1$. This validates the effectiveness of the incremental interpolation scheme in our method.
A qualitative comparison is provided in Fig. \ref{fig:inner}. As can be observed, setting $N=2$ induces an interpolation model capable of producing more accurate structures and textures.

\vspace{1mm}
\noindent \textbf{Efficacy of Memory Bank.} The adoption of the memory bank, which is used for storing common patterns. 
If the memory mechanism is not applied in the final variant of our method, the reduction on the PSNR metric reaches 0.63dB. 

\vspace{1mm}
\noindent \textbf{Performance of Using Different Ensemble Strategies.}
We report the results of single-view models and their simple combinations in Table \ref{tab:ensemble}. `SInt-A', `PInt-C', and `PInt-S' stands for slice-wise interpolation in axial view, pixel-wise interpolation in coronal view, and pixel-wise interpolation in sagittal view, respectively. `PInt-C/PInt-S+SInt-A' indicates `PInt-C' or `PInt-S' is integrated with `SInt-A' through averaging their predictions. `PInt-C+PInt-S+SInt-A' average the predictions of the three models. The simple combinations of pixel-wise and slice-wise interpolation models can improve the results of single models, which demonstrates that the two kinds of models are complementary to each other. Meanwhile, our proposed cross-view mutual distillation can help the combination strategies achieve much better performance.

\vspace{1mm}
\noindent \textbf{Efficacy of Constraint on Wavelet Coefficients.}
The constraint on the wavelet coefficients emphasizes the reconstruction of high-frequency information. Without using the constraint on the wavelet coefficients, the PSNR metric is reduced by 0.55dB.

\input{figures/fig-gamma}
\vspace{1mm}
\noindent \textbf{Using Different Values for $\gamma$.}
In Fig.~\ref{fig:gamma}, we discuss the impact of using different values for the parameter $\gamma$, namely the percents of points used for calculating consistency losses (\ref{eq:loss-cc-ac}) and (\ref{eq:loss-cc-as}). When the deviation between the inferences of slice-wise and pixel-wise interpolation models is too large, one of the two models must predict an incorrect output. However, it is unable to identify which model is more reliable. Hence, we neglect those points at which the loss values are too large. From Fig.~\ref{fig:gamma}, we can see that our method achieves the best performance when $\gamma=25\%$. 


%% file: figures/fig-comp.tex
\begin{figure*}[t]
\centering
\includegraphics[width=0.95\linewidth]{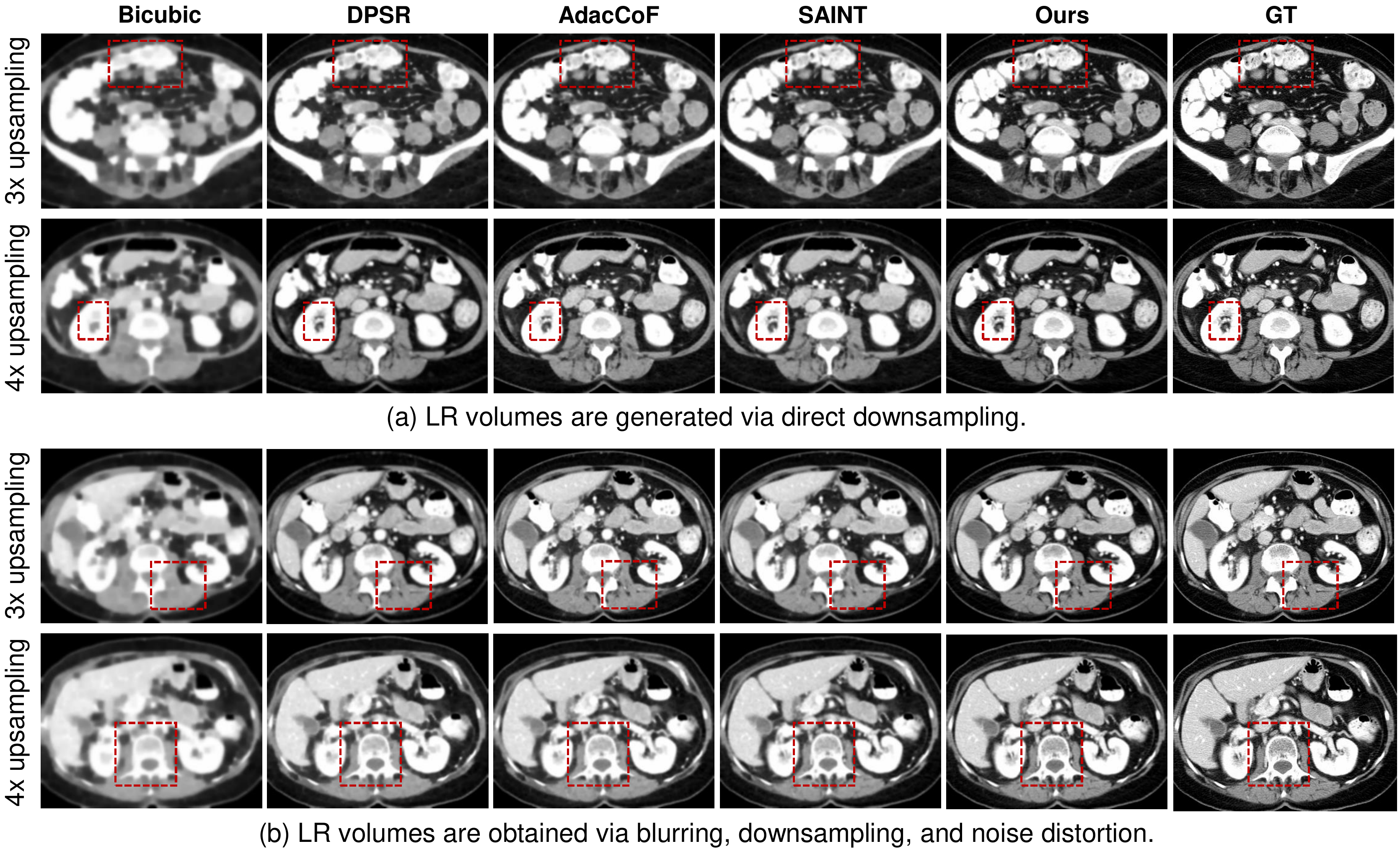}\vspace{-3mm}
\caption{ Qualitative comparisons against existing slice synthesis algorithms on CT-s and MRI-s. The slices synthesized by our method are better than the results of DPSR~\cite{zhang2019deep}, AdaCoF~\cite{lee2020adacof}, and SAINT~\cite{peng2020saint}. (Best viewed in close-up)}\label{fig:comp}\vspace{-5mm}
\end{figure*}

%% file: tables/tab-ct.tex
\begin{table*}[t]
\caption{Comparison with existing slice synthesis, pixel-wise interpolation, and slice-wise interpolation algorithms on the CT dataset, under $2\times$, $3\times$, and $4\times$ upsampling settings. LR volumes are generated via direct downsampling.}\vspace{-3mm}
\label{tab:ct}
\centering
\footnotesize
\setlength{\tabcolsep}{2mm}{
    \begin{tabular}{l|c|c|c|c| c|c|c|c |c|c|c|c}
        \toprule
        \multirow{2}{*}{Method} & \multicolumn{4}{c|}{2$\times$} & \multicolumn{4}{c|}{3$\times$} & \multicolumn{4}{c}{4$\times$} \\ \cmidrule(l){2-5} \cmidrule(l){6-9} \cmidrule(l){10-13}
          & PSNR & SSIM$_a$ & SSIM$_c$ & SSIM$_s$ & PSNR & SSIM$_a$ & SSIM$_c$ & SSIM$_s$ & PSNR & SSIM$_a$ & SSIM$_c$ & SSIM$_s$ \\ \midrule
        RDN~\cite{zhang2018residual}    & 43.51 &0.9539 &0.9519 &0.9512 &39.52 &0.9402 &0.9398 &0.9376 &37.89 &0.9199 &0.9210 &0.9212  \\
        DPSR~\cite{zhang2019deep} & 43.83 &0.9690 &0.9691 &0.9682 &38.82 &0.9434 &0.9423 &0.9424 &38.13 &0.9166 &0.9135 &0.9154  \\
        MetaSR~\cite{hu2019meta} & 43.68 & 0.9547 & 0.9549 & 0.9548 & 39.90 & 0.9419 & 0.9425 & 0.9414 &38.00 &0.9211 &0.9198 &0.9214  \\
        RRIN~\cite{li2020video} & 43.45 &0.9688 &0.9691 &0.9682 & 38.68 & 0.9428 & 0.9424 & 0.9422 & 38.10 & 0.9255 & 0.9232 & 0.9252  \\
        SRGAN~\cite{ledig2017photo} & 43.22 & 0.9524 & 0.9521 & 0.9522 & 38.54 & 0.9433 & 0.9429 & 0.9425 & 37.91 & 0.9213 & 0.9209 & 0.9207  \\
        3D-MDSR~\cite{lim2017enhanced}               & 44.31 & 0.9692 & 0.9698 & 0.9689 & 40.22 & 0.9489 & 0.9489 & 0.9490 & 38.20 & 0.9307 & 0.9302 & 0.9310  \\
        AdaCoF~\cite{lee2020adacof}                    & 44.88 & 0.9749 & 0.9746 & 0.9747 & 40.92 & 0.9513 & 0.9498 & 0.9451 & 38.23 & 0.9311 & 0.9148 & 0.9150 \\ 
        SAINT~\cite{peng2020saint}                     & 44.43 & 0.9694 & 0.9641 & 0.9632 & 40.81 & 0.9448 & 0.9388 & 0.9416 & 38.42 & 0.9259 & 0.9175 & 0.9203  \\ \midrule
        Ours       & \textbf{46.81} &\textbf{0.9792} &\textbf{0.9784} &\textbf{0.9786} &\textbf{42.94} &\textbf{0.9631} &\textbf{0.9589} &\textbf{0.9604} &\textbf{41.11} &\textbf{0.9404} &\textbf{0.9385} &\textbf{0.9382}  \\
        \bottomrule
    \end{tabular} %
}
\end{table*}

%% file: tables/tab-ct-blurred.tex
\begin{table*}[t]
\caption{Comparison with existing slice synthesis, pixel-wise interpolation, and slice-wise interpolation algorithms on the CT dataset, under $2\times$, $3\times$, and $4\times$ upsampling settings. LR volumes are generated via blurring, downsampling and noise distortion.} \vspace{-3mm}
\label{tab:ct-blurred}
\centering
\footnotesize
\setlength{\tabcolsep}{2mm}{
    \begin{tabular}{l|c|c|c|c| c|c|c|c |c|c|c|c}
        \toprule
        \multirow{2}{*}{Method} & \multicolumn{4}{c|}{2$\times$} & \multicolumn{4}{c|}{3$\times$} & \multicolumn{4}{c}{4$\times$} \\ \cmidrule(l){2-5} \cmidrule(l){6-9} \cmidrule(l){10-13}
          & PSNR & SSIM$_a$ & SSIM$_c$ & SSIM$_s$ & PSNR & SSIM$_a$ & SSIM$_c$ & SSIM$_s$ & PSNR & SSIM$_a$ & SSIM$_c$ & SSIM$_s$ \\ \midrule
        RDN~\cite{zhang2018residual} & 41.67 &0.9366 &0.9369 &0.9373 &37.24 &0.9210 &0.9214 &0.9211 &35.23 &0.9004 &0.9010 &0.9011  \\
        DPSR~\cite{zhang2019deep} & 41.92 &0.9389 &0.9391 &0.9387 &37.87 &0.9221 &0.9223 &0.9225 &35.98 &0.9022 &0.9025 &0.9021  \\
        MetaSR~\cite{hu2019meta} & 41.99 &0.9392 &0.9398 &0.9390 &37.95 &0.9262 &0.9259 &0.9264 &36.20 &0.9078 &0.9081 &0.9084  \\
        RRIN~\cite{li2020video} & 41.43 &0.9344 &0.9341 &0.9336 &37.35 &0.9234 &0.9226 &0.9233 &35.58 &0.9045 &0.9045 &0.9054  \\
        SRGAN~\cite{ledig2017photo}                    & 41.10 &0.9319 &0.9313 &0.9321 &37.04 &0.9204 &0.9201 &0.9207 &35.09 &0.8992 &0.9004 &0.9001  \\
        3D-MDSR~\cite{lim2017enhanced}                     & 42.03 &0.9411 &0.9406 &0.9412 &38.25 &0.9310 &0.9303 &0.9306 &36.21 &0.9112 &0.9114 &0.9115  \\
        AdaCoF~\cite{lee2020adacof} & 42.36 &0.9439 &0.9436 &0.9427 &38.72 &0.9313 &0.9311 &0.9320 &36.63 &0.9131 &0.9124 &0.9142 \\ 
        SAINT~\cite{peng2020saint}  & 42.43 &0.9434 &0.9431 &0.9432 &38.88 &0.9352 &0.9358 &0.9348 &36.70 &0.9139 &0.9134 &0.9133  \\ \midrule
        Ours       & \textbf{43.98}  & \textbf{0.9570} & \textbf{0.9568} & \textbf{0.9569} & \textbf{40.91} & \textbf{0.9505} & \textbf{0.9499} & \textbf{0.9499} & \textbf{37.87} & \textbf{0.9244} & \textbf{0.9239} & \textbf{0.9248}\\
        \bottomrule
    \end{tabular}
}
\vspace{-3mm}
\end{table*}

%% file: figures/fig-real.tex
\begin{figure}[t]
\centering
\includegraphics[width=0.95\linewidth]{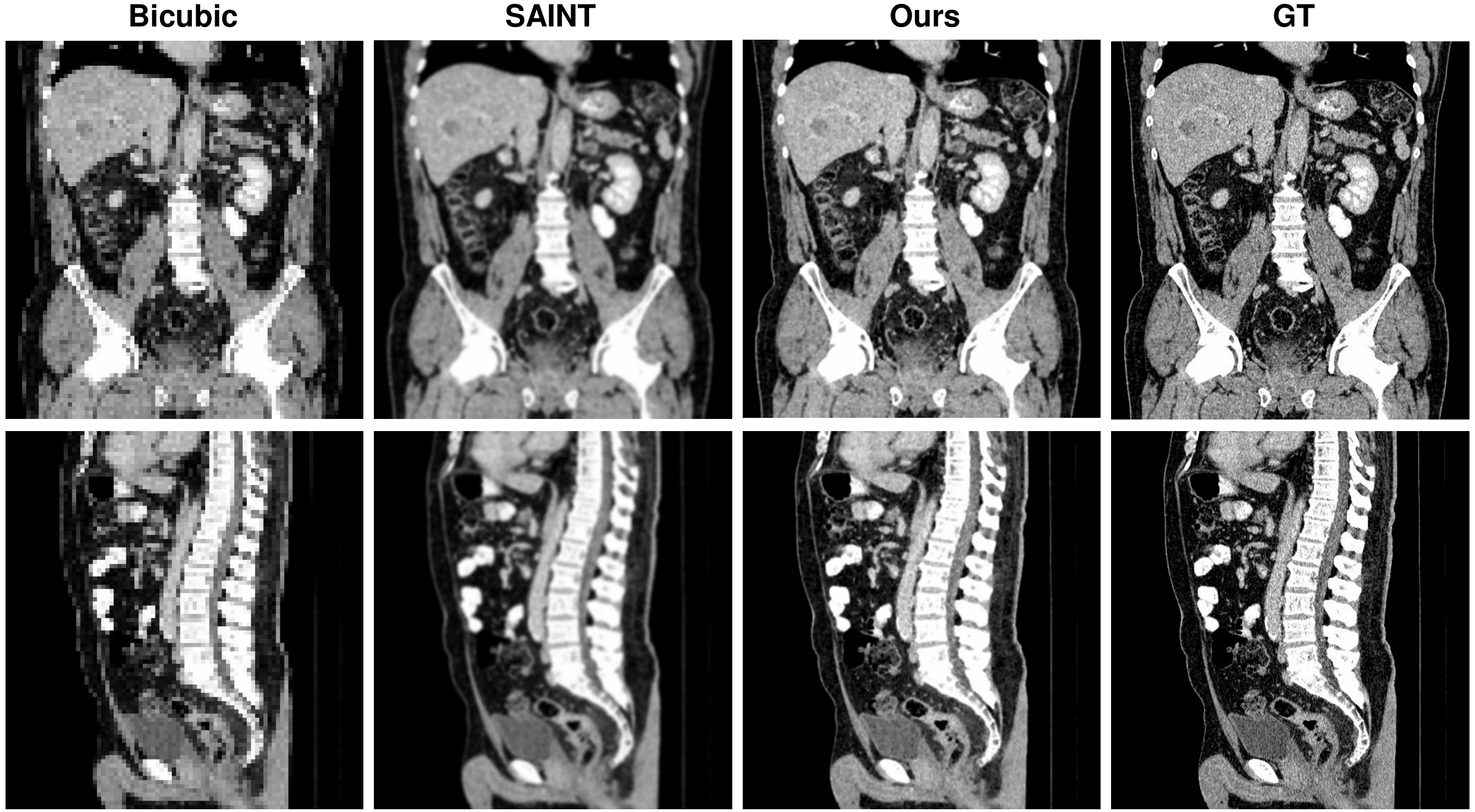} \vspace{-3mm}
\caption{Visualization comparison. From left to right: bicubic interpolation; SAINT; our method, and ground-truth.}\label{fig:real} \vspace{-3mm}
\end{figure}

%% file: tables/tab-ablation.tex
\begin{table}[t]
\caption{Ablation study on critical components in our method.
`w/o $L_c^{cmd}$ or $L_s^{cmd}$' means both $L_c^{cmd}$ and $L_s^{cmd}$ are not used for training. `w/o $L_c^{cmd}$' (`w/o $L_s^{cmd}$') means $L_c^{cmd}$ ($L_s^{cmd}$) is not used. `w/o WT' means the loss on wavelet coefficients is not adopted. `w/o memory' means the memory bank is not applied. For every variant, other parameters are set as in Section~\ref{sec:exper-setting}.
} \vspace{-3mm}
\label{tab:ablation}
\centering
\footnotesize
\setlength{\tabcolsep}{7mm}{
    \begin{tabular}{ l|c|c}
        \toprule
        Variant & PSNR & SSIM$_\textrm{a}$ \\ \midrule
        w/o $L_c^{cmd}$ or $L_s^{cmd}$ & 38.58 & 0.9286 \\ 
        w/o $L_c^{cmd}$ & 40.26 & 0.9322 \\ 
        w/o $L_s^{cmd}$  & 40.24 & 0.9321 \\ 
        N=1           & 40.47 & 0.9325 \\ 
        w/o WT        & 40.56 & 0.9334 \\ 
        w/o memory    & 40.28 & 0.9324 \\ 
        final variant & 41.11 & 0.9404 \\ \bottomrule
    \end{tabular}
}\vspace{-3mm}
\end{table}

%% file: figures/fig-inner.tex
\begin{figure}[t]
\centering
\includegraphics[width=1\linewidth]{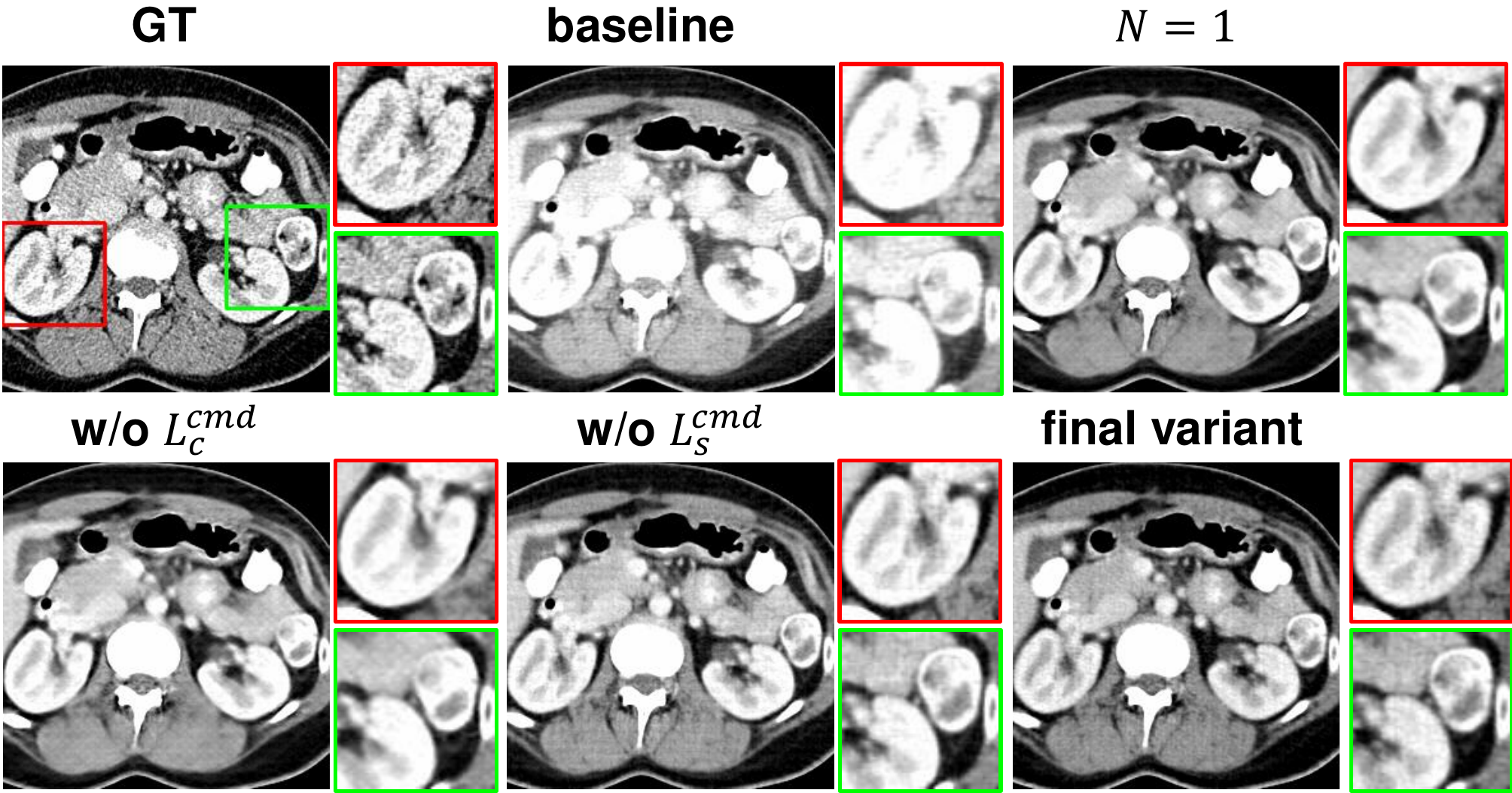} \vspace{-7mm}
\caption{Examples of different variants of our method.}\label{fig:inner}
\vspace{-3mm}
\end{figure}

%% file: tables/tab-ensemble.tex
\begin{table}[t]
\caption{Performance of different ensemble strategies for merging interpolation models.} \vspace{-3mm}
\label{tab:ensemble}
\centering
\footnotesize
\setlength{\tabcolsep}{5mm}{
    \begin{tabular}{l|c|c}
        \toprule
       
        Strategies & PSNR & SSIM$_a$  \\ \midrule
        SInt-A        &38.16 &0.9140  \\
        SInt-A+PInt-C &38.47  &0.9254  \\
        SInt-A+PInt-S &38.49  &0.9261 \\
        SInt-A+PInt-C+PInt-S &38.58  &0.9286 \\ \midrule
        Ours SInt-A   &38.49 &0.9327 \\
        Ours SInt-A+PInt-C   & 40.24 & 0.9355 \\
        Ours SInt-A+PInt-S   & 40.26 & 0.9347 \\
        Ours SInt-A+PInt-C+PInt-S  &\textbf{41.11} & \textbf{0.9404} \\ 
        \bottomrule
    \end{tabular} \vspace{-5mm}
}
\end{table}

%% file: figures/fig-gamma.tex
\begin{figure}[t]
\centering
\includegraphics[width=0.49\linewidth]{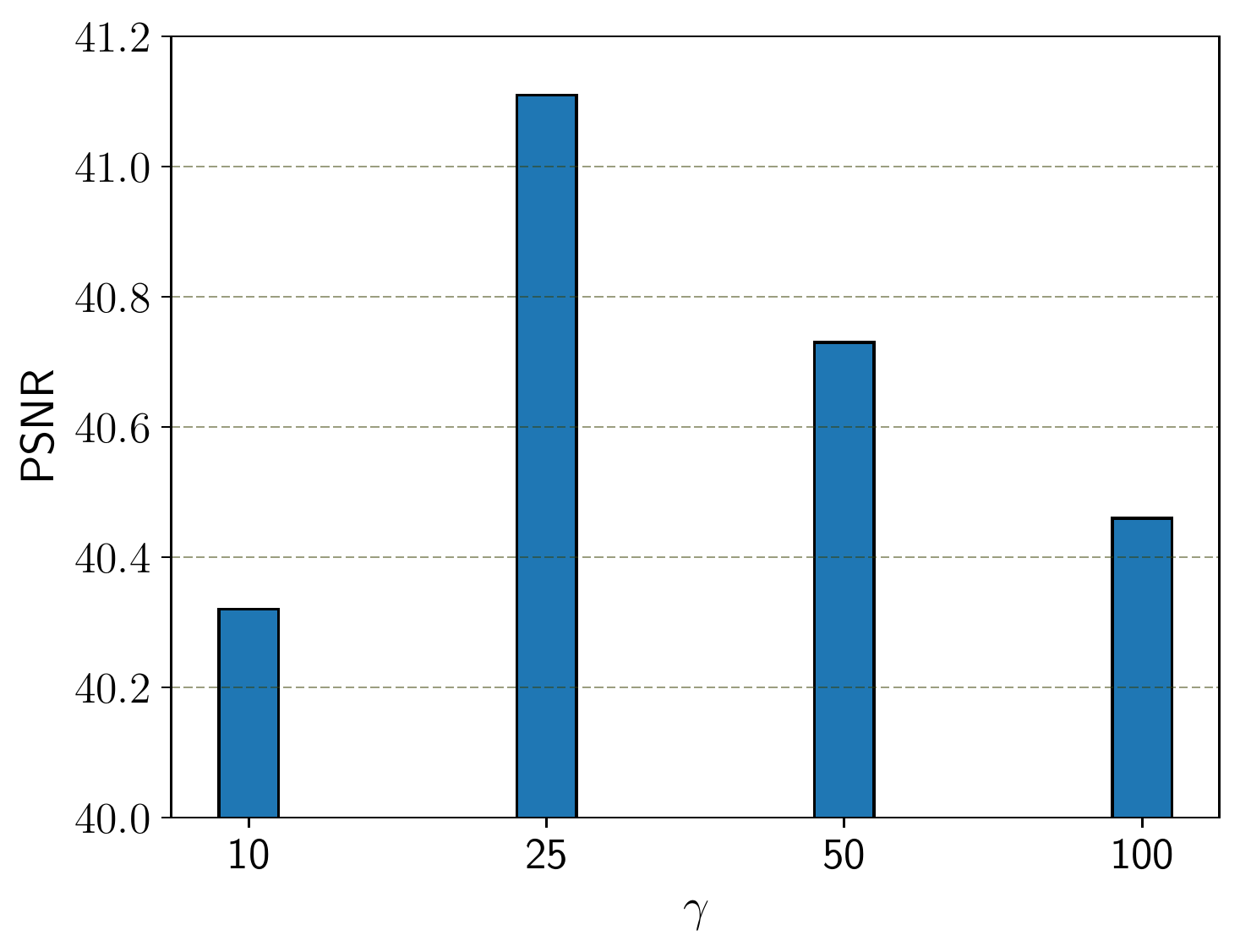}
\includegraphics[width=0.49\linewidth]{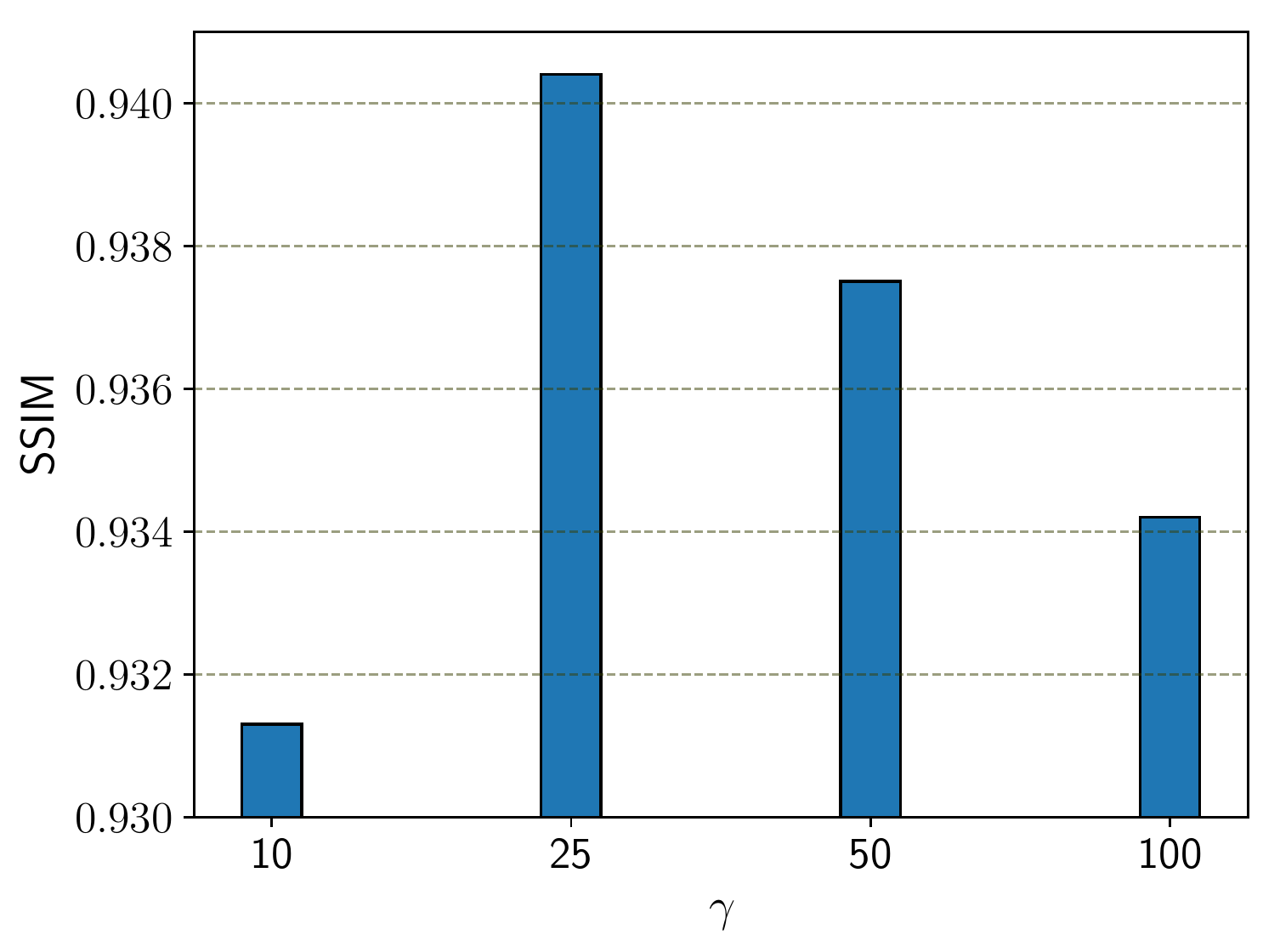}
\vspace{-3mm}
\caption{Performance of using different values for $\gamma$ (\%). 
}\label{fig:gamma}
\vspace{-5mm}
\end{figure}

%% file: sections/conclusion.tex
\section{Conclusions}
This paper proposes an incremental cross-view mutual distillation pipeline to tackle the self-supervised slice synthesis task. The mutual distillation between the slice-wise interpolation in the axial view and pixel-wise interpolation in the coronal and sagittal views contributes to a slice synthesizer with appealing performance. The learning process can be further enhanced via incrementally interpolating intermediate slices and then imposing cross-view distillation on these finer and finer intermediate slices. Extensive experiments on the CT dataset demonstrate the superiority of our method against existing slice synthesis methods.
\vspace{1mm}

\noindent \textbf{Broader Impacts.} Slices synthesized by our method still have apparent difference to real slices. In clinical applications, there exist risks for misleading the disease diagnosis process. It requires further research to improve the practicality of our method.

\vspace{1mm}
\noindent \textbf{Limitations.} In practical clinical scene, there exist many complicated artifacts during the acquisition of LR volumes, such as partial volume effect, motion blur, and streaks. In the current internal learning of our method, we use a simple way to approximate these artifacts. In the future, it deserves in-depth research on modeling the generation of these imaging artifacts for improving the generalization capacity in interpolating real-world LR CTs. 